\documentclass{Interspeech2024}
\usepackage{amsmath,graphicx}
\usepackage{soul}
\usepackage{multirow}
\usepackage{enumerate}
\usepackage{enumitem}
\usepackage{makecell}
\usepackage{flushend}
\usepackage{subcaption}
\usepackage{breqn}

\interspeechcameraready

\hypersetup{
    colorlinks=true,
    linkcolor=blue,
    filecolor=magenta,      
    urlcolor=cyan,
}

\graphicspath{ {Figures/} }

\title{Investigating Confidence Estimation Measures for Speaker Diarization}

\name[]{Anurag}{Chowdhury}
\name[]{Abhinav}{Misra}
\name[]{Mark C.}{Fuhs}
\name[]{Monika}{Woszczyna}
\address{Solventum, USA}
\email{achowdhury2@solventum.com, amisra2@solventum.com, mark.fuhs@solventum.com, mwoszczyna@solventum.com}

\keywords{Speaker Diarization, Confidence Estimation, Speaker Recognition, Speaker Clustering}

\begin{document}

\maketitle

\begin{abstract}
Speaker diarization systems segment a conversation recording based on the speakers' identity. Such systems can misclassify the speaker of a portion of audio due to a variety of factors, such as speech pattern variation, background noise, and overlapping speech. These errors propagate to, and can adversely affect, downstream systems that rely on the speaker's identity, such as speaker-adapted speech recognition. One of the ways to mitigate these errors is to provide segment-level diarization confidence scores to downstream systems. In this work, we investigate multiple methods for generating diarization confidence scores, including those derived from the original diarization system and those derived from an external model. Our experiments across multiple datasets and diarization systems demonstrate that the most competitive confidence score methods can isolate $\sim$30\% of the diarization errors within segments with the lowest $\sim$10\% of confidence scores. 
\end{abstract}

\vspace{-0.2cm}
\section{Introduction}~\label{sec:intro}
Speaker diarization is the task of determining ``who spoke when" in speech audio. Traditionally, we perform speaker diarization by segmenting speech audio into short segments and then clustering them by their perceived speaker identity. However, such an approach inherits the challenges of speaker recognition, such as low inter-speaker variability and high intra-speaker variability~\cite{kinnunen2010overview}, in addition to the challenges of speaker clustering, such as overlapping and imbalanced amount of speech from an unknown number of speakers~\cite{fujita2019end}. While several techniques have been developed to address these challenges, most can only deal with a small subset of the challenges within their limitations. For example, the recently developed End-to-End (E2E) speaker diarization systems combine speech activity detection (SAD), speaker recognition, and clustering into one E2E system~\cite{fujita2019end} and are adept at diarizing overlapping speech segments. However, their performance worsens with an increasing number of speakers~\cite{horiguchi2020end}. On the other hand, system combination techniques~\cite{raj2021dover} are gaining popularity to create an ensemble of several diarization methods that exceed the performance of its constituent methods. However, such ensemble systems are error-prone in portions of the speech audios where the constituent systems do not agree on a decision.

\begin{figure}[t]		
		\centering
		\includegraphics[scale = 0.2]{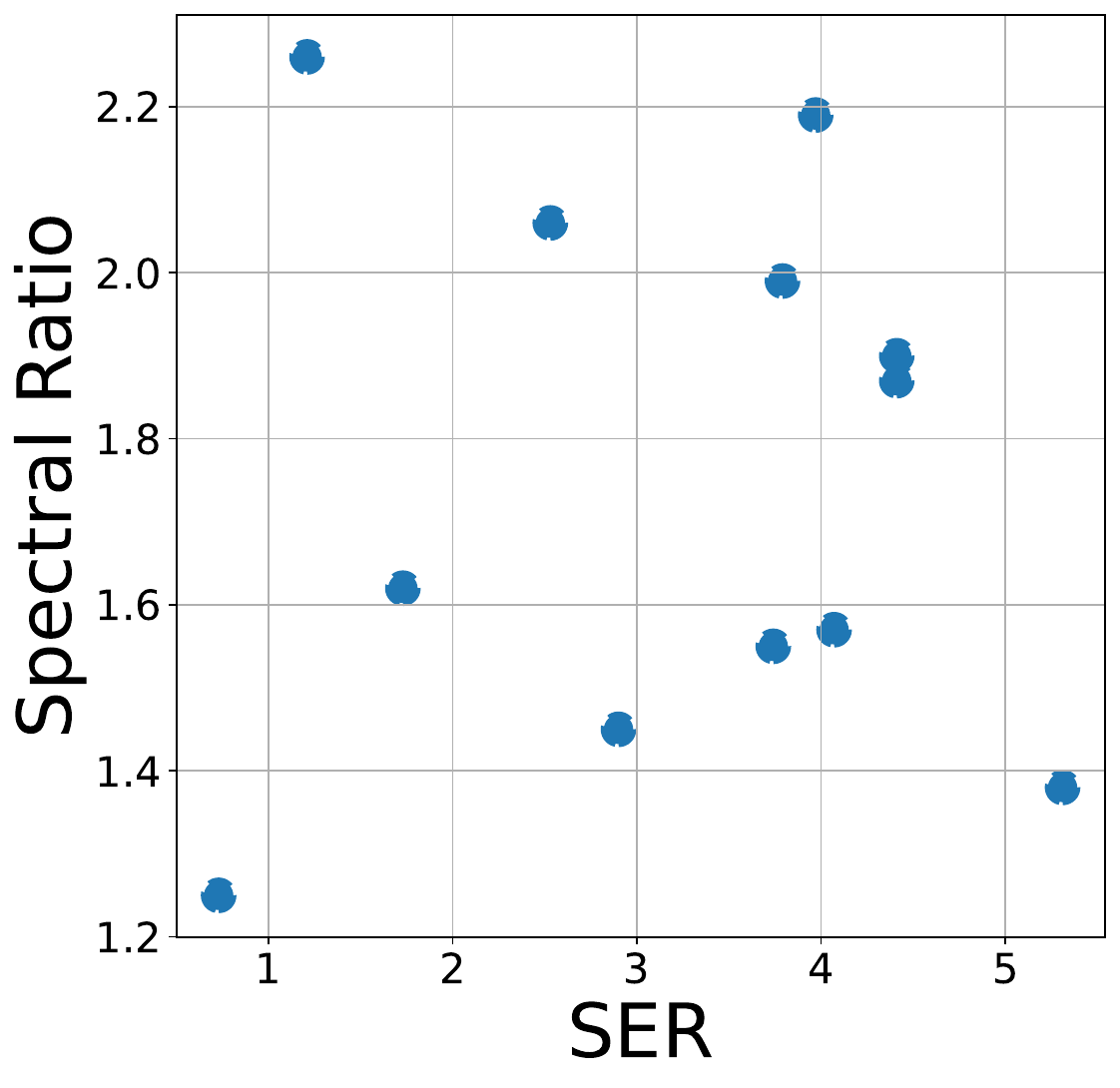}
		\caption{A plot of spectral ratio (SR) vs. speaker error rate (SER) was computed using ECAPA-TDNN speaker embeddings on the eval set of the AMI dataset. No correlation is observed between the SR and SER, indicating that SR is a poor predictor of diarization confidence.
		}
		\label{fig:sr_vs_ser}
\end{figure}

One of the main applications of speaker diarization systems is to segment conversational speech audio into short homogenous speech chunks based on the speaker's identity before feeding them to downstream tasks. For example, automatic conversational transcription systems use speaker diarization together with automatic speech recognition (ASR) to generate turn-by-turn speaker-labeled text transcripts of audio conversations~\cite{park2022review}. However, long conversations often feature nuisance factors, such as low signal-to-noise ratio, background noise, short vocal responses, and overlapping speech, leading to decreased diarization accuracy. 

These diarization inaccuracies can propagate to and compound in downstream tasks. For example, automatic speech recognition systems can benefit from adapting models to individual speakers~\cite{zhao1994acoustic,meng2019speaker}; however, while such models improve accuracy on the target speaker, performance degrades when one speaker's model is applied to a different speaker's speech. For segments where the identity of the speaker is unclear, alternate strategies can be employed, such as using a speaker-independent ASR model or selecting from the output of multiple speaker-adapted models. Similarly, semi-supervised adaptation of an ASR model to a particular speaker depends on identifying speech from that speaker with high precision. Spoken language understanding tasks can also be affected. The statement ``I'd like to try $\langle medication\rangle$ to see if it helps" in a doctor-patient conversation could imply a suggestion if spoken by the patient, or an actionable item in a plan of care if spoken by the doctor. Actionable items may require additional user confirmation where the speaker identity or transcription is low confidence.
 
Existing confidence assessment methods, such as in~\cite{toledo2012confidence}, use the spectral ratio of eigenvalues to estimate conversation-level confidence scores; they do not perform confidence assessment at the segment level. In our experiments on the AMI dataset, we evaluated the effectiveness of spectral ratio as a predictor of diarization performance. As shown in Fig.~\ref{fig:sr_vs_ser}, the spectral ratio of ECAPA-TDNN-speaker embeddings~\cite{desplanques2020ecapa} does not correlate with the Speaker Error Rate (SER) of the diarization system. Spectral ratio, therefore, cannot be used to estimate diarization confidence. 

\begin{figure*}[t]		
		\centering
		\includegraphics[scale = 0.63]{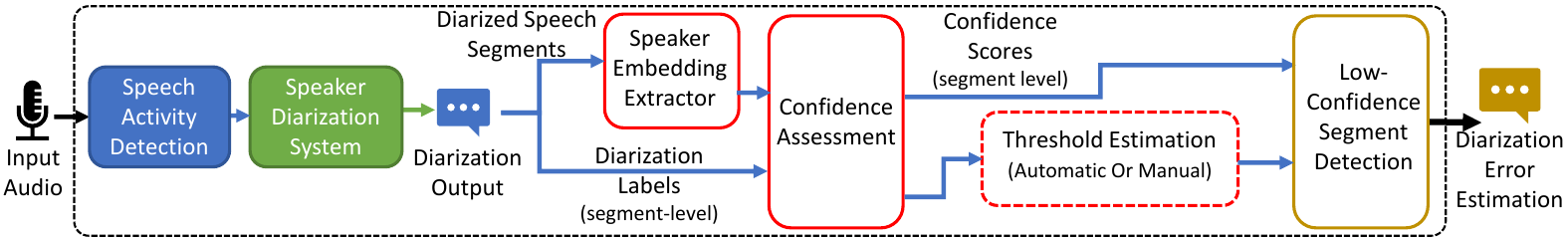}
		\caption{A visual representation of the proposed speaker diarization confidence assessment framework.
		}
		\label{fig:framework}
        \vspace{-0.4cm}
\end{figure*}

In this work, we investigate methods for assigning segment-level diarization confidence scores to several different types of diarization systems, including xVector- and ECAPA-TDNN-based systems using spectral clustering, an end-to-end diarization system, and output from a multi-system combination. We consider both white-box and black-box scenarios, where confidence is assigned either based on the model of the original diarization system or using a secondary system, the latter being particularly important in the multi-system case.

To analyze the effectiveness of the proposed confidence assessment method, we study the proportion of incorrectly diarized segments isolated in the low-confidence segments. In our experiments, across multiple diarization systems and datasets, we isolated almost 30\% of the diarization errors in the 10\% of segments with the lowest confidence. This demonstrates the proposed method's efficacy in identifying incorrect diarization outputs without prior knowledge or access to the diarization system. We then explore the distributions of speaker embeddings and their confidence scores to add insight into confidence score performance.

\vspace{-0.2cm}
\section{Methods}~\label{sec:ProposedMethod}
For all of the diarization systems we consider, conversational audio is first segmented into continuous speech segments by a speech activity detector (SAD). The SAD model comprises five TDNN layers interleaved with two LSTM layers, and speech / non-speech output posteriors are smoothed with median filtering~\cite{VAD_TDNN_LSTM}. We then apply one (or combination of) diarization systems to the speech portions to segment them into single-speaker segments and label each segment with a speaker index. Confidence scores are then computed for each segment using each relevant method.

\subsection{Confidence Assessment}~\label{sec:ConfAss}
In this work, we investigate the following different methods for computing the confidence scores. In each case, we rely on a speaker embedding per segment that is generated by the original diarization system's model or by a secondary system. Speaker centroids are the average of the embeddings associated with all of the segments assigned to a speaker. 
\begin{itemize}[leftmargin=0cm,itemindent=.5cm,labelwidth=\itemindent,labelsep=0cm,align=left]
\setlength\itemsep{-0.3em}

\item \textit{Cosine Similarity Score}: This method computes confidence scores as the mean cosine similarity between a segment's speaker embeddings and the speaker's centroid. 

\item \textit{Local Confidence Score}: This method uses the cosine similarity of speech segment embeddings to their predicted speaker centroids to compute the initial confidence scores. Next, we drop the speech embeddings two standard deviations away from the centroid for each speaker cluster and use the remaining embeddings in each cluster to re-estimate their centroids. We repeat this step until the centroid stabilizes for each speaker cluster. We then use the cosine similarity of the updated speaker centroids with their cluster members to compute the final confidence scores.

\item \textit{Silhouette Score}: This method uses the silhouette score~\cite{shahapure2020cluster}, a clustering validation metric to compare embeddings with the predicted speaker and also the closest other speaker centroid:
\vspace{-0.3cm}
\begin{dmath}
s(x) = \frac{b(x)-a(x)}{max(a(x),b(x))} 
\end{dmath}
\vspace{-0.3cm}
Here, $s(x)$ is the silhouette score of a speech embedding $x$. $b(x)$ is the cosine distance of $x$ to the centroid of the closest cluster that it is not a part of and $a(x)$ is the cosine distance of $x$ to the centroid of its own cluster. Silhouette score computation needs at least two clusters; therefore, we revert to using the cosine similarity metric for single-speaker audios (such as dictations of notes) as our confidence measure.

\item \textit{Spectral Clustering Score}: A variant of the Cosine Similarity Score using the eigenbasis from spectral clustering.

\end{itemize}

\subsection{Spectral Clustering}~\label{sec:SpecClust}
Briefly, spectral clustering is a three-step process. First, given $N$ $D$-dim embeddings, the $N$x$D$ embedding matrix is unit-normalized and the outer product is computed to create an $N$x$N$ cross-correlation matrix. Then, eigendecomposition of the cross-correlation matrix reveals $S$ large and many small eigenvalues, separated by a so-called eigen-gap~\cite{von2007tutorial}, corresponding to $S$ different speakers in the conversation; the smaller eigenvalues are discarded. Each large eigenvalue's $N$-dim eigenvector encodes the embeddings for one speaker with higher values, whereas embeddings for other speakers are near 0. Finally, the cross-correlation vector for each embedding is then represented as an $S$-dim basis vector over the $S$ retained eigenvectors, and the argmax of this vector indicates the speaker to which the embedding is assigned.

A spectral variant of cosine similarity can be derived using the spectral basis.
Each speaker's centroid is estimated from the $S$-dimensional embedding basis vectors assigned to that speaker. Confidence scores are then computed as the cosine similarity between each embedding and it's assigned speaker centroid, both in the spectral basis. Where different portions of a segment are covered by different embeddings, the cosine similarity scores are averaged.

\subsection{Confidence Estimation Metrics}~\label{sec:ConfMetrix}
Confidence scores are evaluated in a rank-order context: we partition the segments of each conversation into low- and high-confidence subsets, containing the segments with the lowest (1-Cov)\% and highest Cov\% of confidence scores, respectively. The following metrics, computed on the high-confidence subset, are then reported:
\begin{itemize}[leftmargin=0cm,itemindent=.5cm,labelwidth=\itemindent,labelsep=0cm,align=left]
	\setlength\itemsep{0em}
    \item Coverage (Cov): This metric indicates the proportion of the diarized audio length in the entire conversation that attains a high-confidence value. This is an independent variable that we evaluate at two operating points: 70\% coverage and 90\% coverage.
    
    \item Covered Diarization Error Rate (cDER): This metric estimates the diarization error rate (DER) within the covered region of the diarization output. In order to prevent double counting the false negatives, the low-confidence speech segments are reported by the coverage metric and are not counted as a `Miss.' If confidence scores were random, then cDER would approximate DER. Informative confidence scores should shift high-error segments out of the high-confidence subset, lowering the cDER.
\end{itemize}

\subsection{Global Thresholding for Data Selection}~\label{sec:GlobThresh}
To directly compare cDERs at the same operating point, we evaluated at fixed coverage percentages. However, for down-stream tasks relying on semi-supervised data selection, one may wish to automatically select a subset of the data that maximizes the amount of data selected while minimizing the DER of the selected data (cDER). Here, a validation set is employed to find the minimum of the ratio of cDER to coverage, which varied between 50\% and 90\% coverage on our datasets. The exact minimization criterion could be adjusted depending on the sensitivity of the down-stream task to precision vs. recall errors.

\begin{table}[ht]
	\fontsize{6}{8}\selectfont
	\caption{Diarization confidence assessment results on the DoPaCo and AMI datasets, given at fixed coverage values (Cov.) of 70\% and 90\%. The best performing confidence assessment method (T.M.) for each diarization method (D.M.) is marked in bold. Note the results reported in bold for the T1 method (no thresholding) are given at a coverage of 100\%. The T2 method gives baseline (spectral clustering-based) results. The T3 to T5 methods give results of the proposed confidence assessment methods}
	\centering	
	 
    \vspace{-0.1cm}
    \scalebox{1.1}{
    \centering
    

\begin{tabular}{|p{2.5mm}|p{2mm}|c|c|c|c|}
\hline 
\multicolumn{2}{|c|}{Dataset} & \multicolumn{2}{c|}{AMI Dataset} & \multicolumn{2}{c|}{Dopaco Dataset}\tabularnewline
\hline 
\multirow{2}{2.5mm}{D.M.} & \multirow{2}{2mm}{T.M.} & \multicolumn{1}{c|}{Cov. = 70\%} & \multicolumn{1}{c|}{Cov. = 90\%} & \multicolumn{1}{c|}{Cov. = 70\%} & \multicolumn{1}{c|}{Cov. = 90\%}\tabularnewline
\cline{3-6} \cline{4-6} \cline{5-6} \cline{6-6} 
 &  & cDER  & cDER  & cDER  & cDER\tabularnewline
\hline 
\multirow{5}{2.5mm}{M1} & T1  & \multicolumn{2}{c|}{\textit{12.16 }} & \multicolumn{2}{c|}{\textit{17.48}}\tabularnewline
\cline{2-6} \cline{3-6} \cline{4-6} \cline{5-6} \cline{6-6} 
 & T2  & 6.19  & 10.19  & 12.59  & 15.66\tabularnewline
\cline{2-6} \cline{3-6} \cline{4-6} \cline{5-6} \cline{6-6} 
 & T3  & 3.85  & 8.8  & 8.06  & 14.27\tabularnewline
\cline{2-6} \cline{3-6} \cline{4-6} \cline{5-6} \cline{6-6} 
 & T4  & 4.71  & 7.75  & 7.98  & 13.49\tabularnewline
\cline{2-6} \cline{3-6} \cline{4-6} \cline{5-6} \cline{6-6} 
 & T5  & \textbf{3.67}  & \textbf{6.66}  & \textbf{7.78}  & \textbf{12.79}\tabularnewline
\hline 
\multirow{5}{2.5mm}{M2} & T1  & \multicolumn{2}{c|}{\textit{5.18 }} & \multicolumn{2}{c|}{\textit{13.03}}\tabularnewline
\cline{2-6} \cline{3-6} \cline{4-6} \cline{5-6} \cline{6-6} 
 & T2  & \textbf{3.33}  & 4.17  & 7.39  & 10.39\tabularnewline
\cline{2-6} \cline{3-6} \cline{4-6} \cline{5-6} \cline{6-6} 
 & T3  & 3.35  & 4.55  & 8.37  & 11.48\tabularnewline
\cline{2-6} \cline{3-6} \cline{4-6} \cline{5-6} \cline{6-6} 
 & T4  & \textbf{3.33}  & 4.17  & 7.39  & 10.39\tabularnewline
\cline{2-6} \cline{3-6} \cline{4-6} \cline{5-6} \cline{6-6} 
 & T5  & 3.34  & \textbf{4.01}  & \textbf{6.67}  & \textbf{9.25}\tabularnewline
\cline{2-6} \cline{3-6} \cline{4-6} \cline{5-6} \cline{6-6} 
 & T6 & 4.31 & 4.89 & N.A & N.A\tabularnewline
\hline 
\multirow{5}{2.5mm}{M3} & T1  & \multicolumn{2}{c|}{\textit{6.9 }} & \multicolumn{2}{c|}{\textit{8.36}}\tabularnewline
\cline{2-6} \cline{3-6} \cline{4-6} \cline{5-6} \cline{6-6} 
 & T2  & 5.28  & 6.05  & 5.51  & 6.98\tabularnewline
\cline{2-6} \cline{3-6} \cline{4-6} \cline{5-6} \cline{6-6} 
 & T3  & 3.7  & 5.37  & \textbf{4.45}  & 6.39\tabularnewline
\cline{2-6} \cline{3-6} \cline{4-6} \cline{5-6} \cline{6-6} 
 & T4  & 3.68  & 4.76  & 4.88  & 6.39\tabularnewline
\cline{2-6} \cline{3-6} \cline{4-6} \cline{5-6} \cline{6-6} 
 & T5  & \textbf{3.67}  & \textbf{4.43}  & 4.63  & \textbf{6.07}\tabularnewline
\hline 
\multirow{5}{2.5mm}{M4} & T1  & \multicolumn{2}{c|}{\textit{3.33}} & \multicolumn{2}{c|}{\textit{8.85}}\tabularnewline
\cline{2-6} \cline{3-6} \cline{4-6} \cline{5-6} \cline{6-6} 
 & T2  & 1.26  & 2.53  & 3.92  & \textbf{7.52}\tabularnewline
\cline{2-6} \cline{3-6} \cline{4-6} \cline{5-6} \cline{6-6} 
 & T3  & 2.51  & 3.06  & 6.51  & 8.08\tabularnewline
\cline{2-6} \cline{3-6} \cline{4-6} \cline{5-6} \cline{6-6} 
 & T4  & 1.26  & 2.53  & 3.92  & \textbf{7.52}\tabularnewline
\cline{2-6} \cline{3-6} \cline{4-6} \cline{5-6} \cline{6-6} 
 & T5  & \textbf{1.2}  & \textbf{2.43}  & \textbf{3.71}  & \textbf{7.52}\tabularnewline
\hline 
\end{tabular}

}

\vspace{0.2cm}
\centering
\begin{tabular}{|c|c|c|c|}
\hline 
M1 & M2 & M3 & M4\tabularnewline
\hline 
\hline 
xVector +SC  & PyAnnote2.0 (E2E) & ECAPA-TDNN +SC & \makecell{DOVER-Lap\\(of M1, M2, M3)}\tabularnewline
\hline 
\end{tabular}
\vspace{0.2cm}
\centering
\scalebox{0.94}{
\begin{tabular}{|c|c|c|c|c|c|}
\hline 
T1 & T2 & T3 & T4 & T5 & T6\tabularnewline
\hline 
\hline 
\makecell{100\%\\Coverage} & \makecell{Spectral\\Clustering\\Score} & \makecell{Local\\Confidence\\Score} & \makecell{Cosine\\Similarity\\Score} & \makecell{Silhouette\\Score} & \makecell{PyAnnote2.0\\(E2E) Score}\tabularnewline
\hline 
\end{tabular}
}

\vspace{-0.5cm}
\label{tab:results}
\end{table}	

\section{Experiments}

\subsection{Datasets}~\label{sec:datasets}
We perform our speaker diarization experiments on the following multi-speaker conversation datasets.
\begin{itemize}[leftmargin=0cm,itemindent=.5cm,labelwidth=\itemindent,labelsep=0cm,align=left]
	\setlength\itemsep{0em}
	\item\textbf{AMI:} This is the Augmented Multi-party Interaction (AMI) meeting dataset~\cite{kraaij2005ami}. We use the official ``Full ASR corpus" split with TNO meetings excluded from the Dev and Eval set. This dataset provides a point of reference for comparing the baseline diarization methods' performance and the corresponding metrics for confidence scores.
	\item\textbf{DoPaCo:} This internal dataset consists of manually de-identified doctor-patient conversations recorded using near-and far-field microphones in semi-unconstrained indoor settings of doctors' examination rooms. We use this dataset to demonstrate the diarization performance of several publicly available state-of-the-art methods on the challenges offered by a doctor-patient conversation.	
\end{itemize}

\subsection{Speaker Diarization Systems}~\label{sec:spk_dia_exp}
We use the SpeechBrain~\cite{ravanelli2021speechbrain} toolkit's speaker diarization setup using the ECAPA-TDNN~\cite{dawalatabad2021ecapa} and xVector~\cite{snyder2018x} models (pretrained on the VoxCeleb~\cite{nagrani2020voxceleb} dataset) paired with spectral clustering (SC) as our first and second speaker diarization systems. The ECAPA-TDNN and xVector-based diarization systems generate embeddings from overlapping windows of length 1.5 secs, shifted by 250ms. PyAnnote 2.0 toolkit's~\cite{Bredin2020} pretrained E2E speaker diarization system~\cite{Bredin2021} is our third diarization system. We also use the DOVER-Lap~\cite{raj2021dover} method to combine the diarization outputs of the above three diarization systems, thereby instituting a fourth diarization system. 
Confidence scores for all methods except the E2E Activation Score were computed using embeddings generated by the ECAPA-TDNN system, since it was overall the strongest. DER and cDER are computed using a collar of 250 ms and exclude overlapping speech segments to maintain consistency with other published works.


\section{Results}~\label{sec:Results}
\begin{figure}
     \centering
     \begin{subfigure}[b]{0.23\textwidth}
         \centering
         \includegraphics[width=\textwidth]{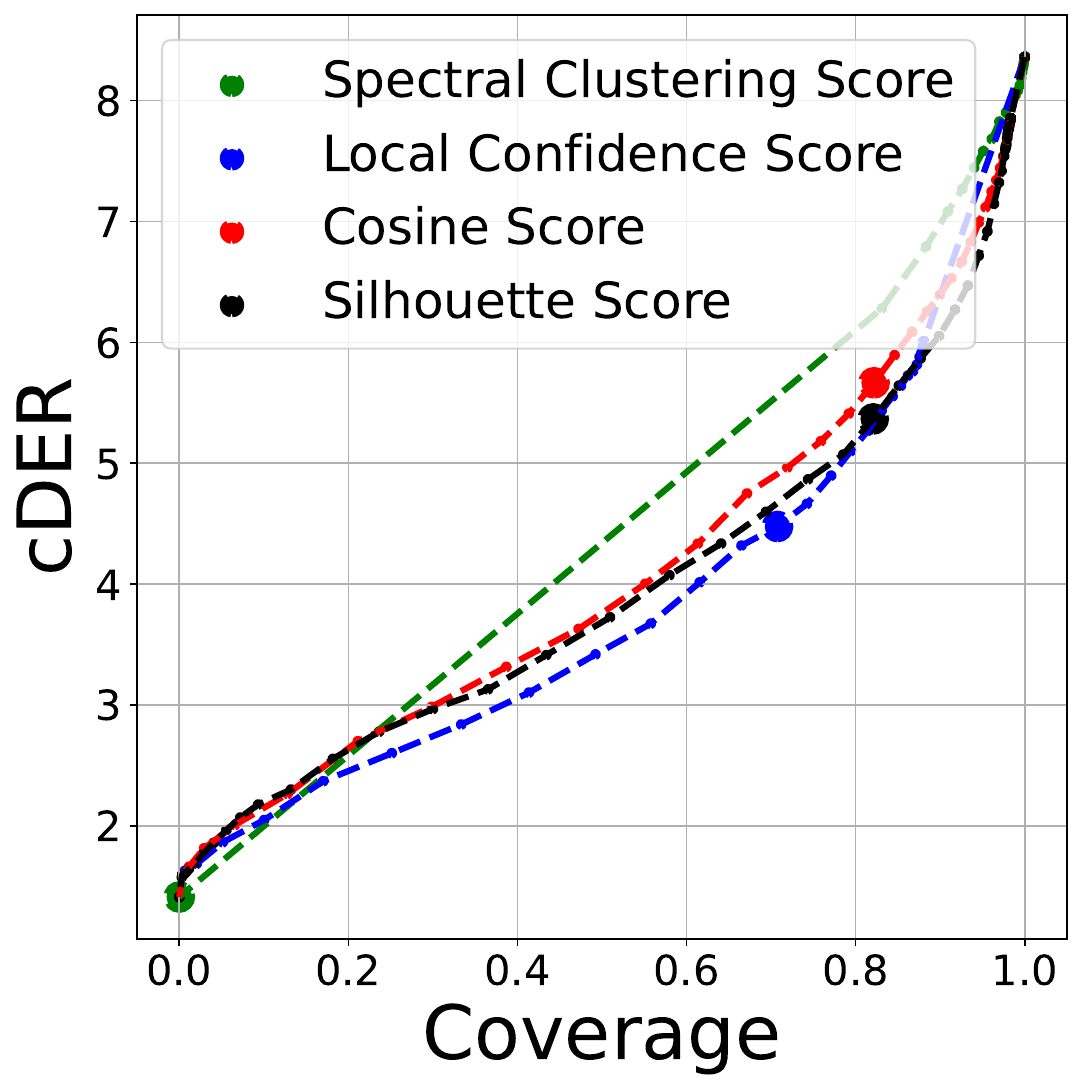}
         \caption{DoPaCo Dataset}
         \label{fig:spec_a}
     \end{subfigure}
     \hfill
     \begin{subfigure}[b]{0.23\textwidth}
         \centering
         \includegraphics[width=\textwidth]{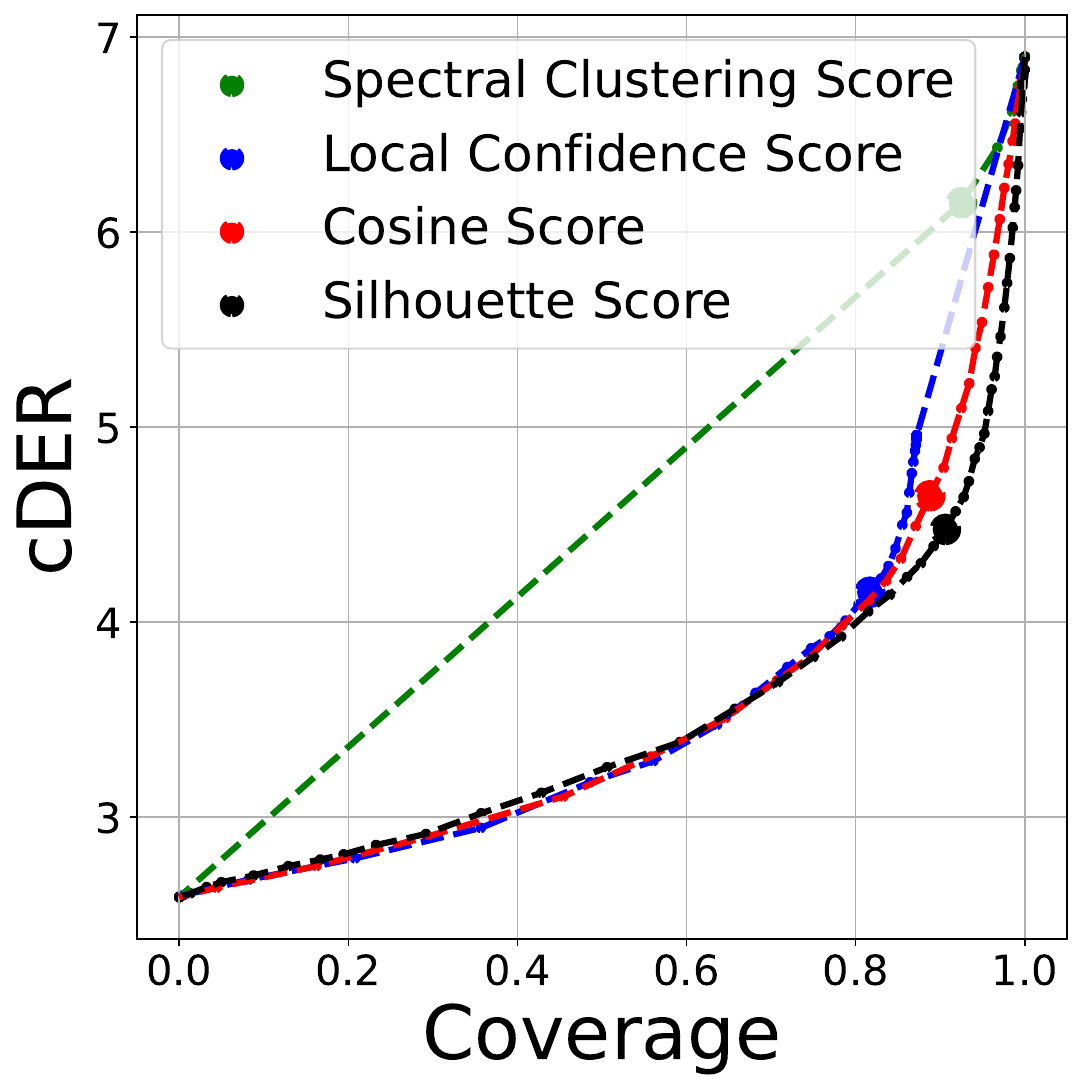}
         \caption{AMI Dataset}
         \label{fig:spec_b}
     \end{subfigure}
     \hfill
        \caption{cDER vs Coverage plots using the ECAPA-TDNN based diarization method. Large markers in the plots show the operating points at global thresholds. }
        \label{fig:cDER_V_coverage_plots}
        \vspace{-0.5cm}
\end{figure}

We report the speaker diarization performance (see Table~\ref{tab:results} and Fig.~\ref{fig:cDER_V_coverage_plots}) using covered Diarization Error Rate (cDER) and coverage metrics. In this set of experiments, we analyze and compare the performance (cDER) of the different confidence assessment methods at fixed coverage values of 70\% and 90\%. 

Overall, silhouette score (T5) was the strongest performer, with the best or a close-second-best cDER across systems and coverages. T3 and T4 were also competitive, whereas using the spectral basis (T2) showed significant degradation in some conditions. On both the datasets and across all the diarization systems, the methods T3, T4, and T5 obtain an average $\sim$55\% and $\sim$31\% reduction in cDER at the fixed coverage values of 70\% and 90\% compared to the overall DER (T1).

M3 is a white-box condition, where the embedding model used for diarization is the same as the model used for confidence score estimation. While we expected that M1 and M2 systems might show more cDER improvement, owing to the use of a different embedding model for confidence scores, this was not a consistent trend. 

T6 is the white-box condition for the E2E system, where embeddings are taken from within and the cosine similarity method is applied. This is in most direct contrast to method T4: cosine similarity over the ECAPA-TDNN embeddings generated on the E2E system's speaker segmentation. T6 was not competitive with the other methods on the AMI dataset, so we did not pursue it further on the DoPaCo dataset.

The DOVER-Lap (M4) system's performance was hampered by the spread of DERs of the base systems. Competitive performance from M2 and M3 on the AMI dataset likely contributed to M4's strong AMI performance. For the DoPaCo dataset, M3 was far ahead of M1 and M2, and the combination did not improve the baseline DER. Interestingly, M4's cDER at 70\% coverage was better than the single M3 system, despite higher DER than M3. Overall, we observe a $\sim$20\% and $\sim$60\% (relative) reduction in cDER at 10\% and 30\% loss of coverage, respectively, using the proposed methods. This demonstrates the efficacy of the methods at identifying low-confidence segments in the combined diarization output of several independent diarization systems.



Figure~\ref{fig:cDER_V_coverage_plots} shows the variation in cDER with coverage values for the proposed and baseline confidence assessment methods. We also mark the operating points corresponding to each technique's optimal threshold obtained by the global thresholding process (Section~\ref{sec:GlobThresh}). Embedding-space methods T3, T4 and T5 show a much larger reduction in cDER values compared to the spectral method (T2) for a similar amount of coverage. For example, on the AMI dataset, methods T3-T5 demonstrate an average relative reduction of 57\% cDER at 80\% coverage. In contrast, the spectral method only reduced the cDER by 19\% at a similar coverage.

\vspace{-0.2cm}
\section{Analysis}~\label{sec:ablation}
\begin{figure}[th!]		
		\centering
		\includegraphics[scale = 0.51]{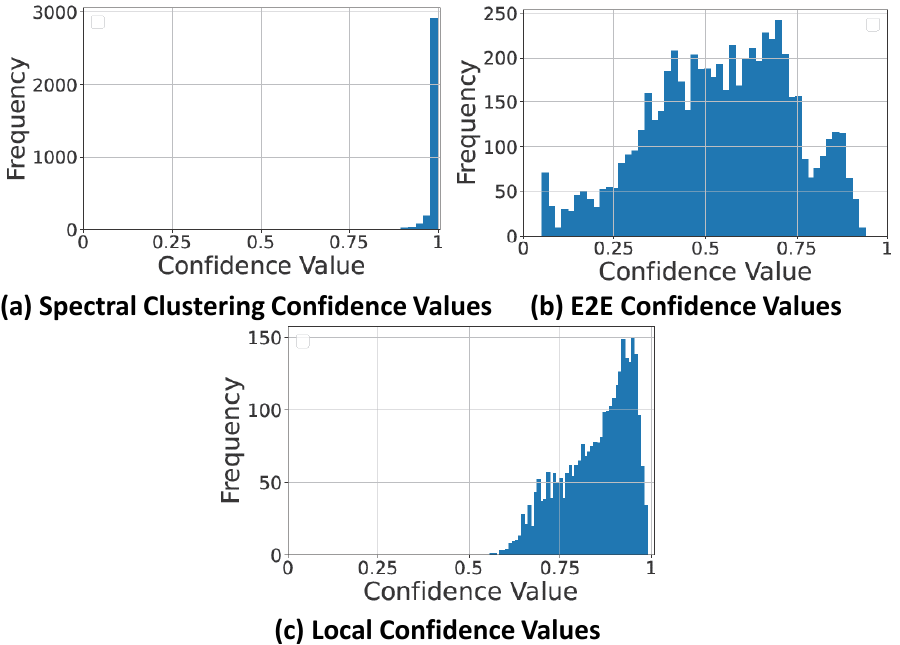}
		\caption{A comparison of histograms of diarization confidences scores estimated using (a) spectral clustering-based, (b) End-to-End-based and (c) proposed confidence assessment methods on the AMI dataset.
		}
		\label{fig:conf_hist}
\end{figure}

The spectral method T2 did not perform as well as the other methods in several cases. An analysis of the distribution of scores generated by T2 revealed a strongly skewed distribution, shown in Fig.~\ref{fig:conf_hist}(a), where there was little range over which to distinguish bad segment labels from the good ones. Scores based on embeddings from the ECAPA-TDNN and E2E models covered a broader range, as shown in Figures~\ref{fig:conf_hist}(b) and ~\ref{fig:conf_hist}(c).

\begin{figure}[t]		
		\centering
		\includegraphics[scale = 0.385]{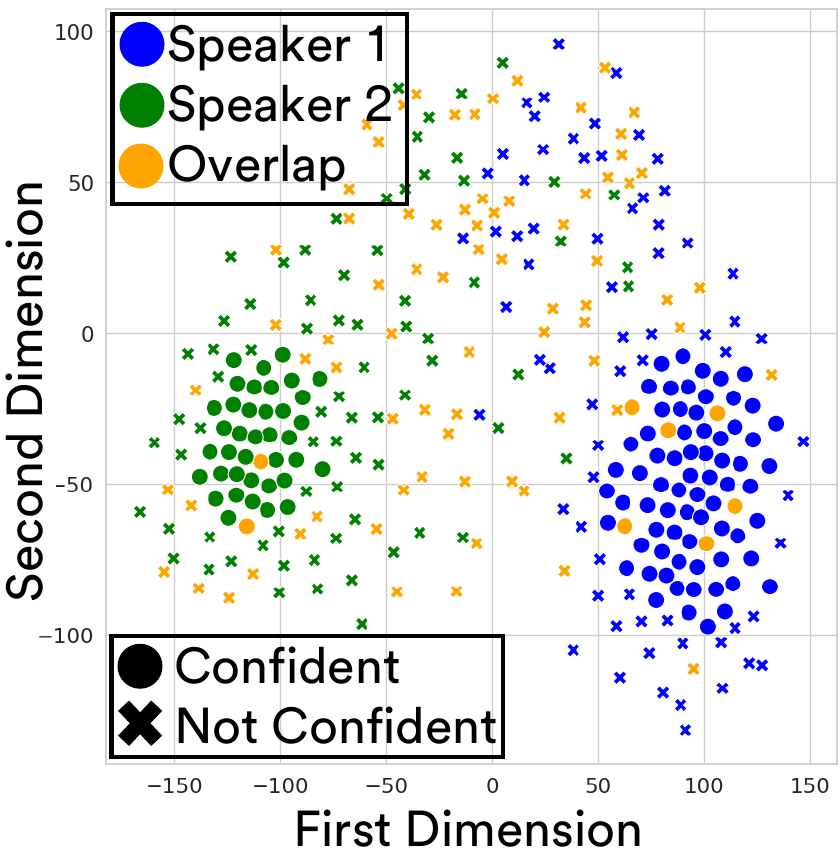}
		\caption{A visual representation of the Local Confidence estimation results on a doctor-patient conversation from the DoPaCo dataset. The proposed method assigns lower confidence to most overlapping speech segments.}
		\label{fig:thresholding}
\end{figure}

To better visualize the distribution of embeddings, Figure~\ref{fig:thresholding} shows the 2-dim t-SNE representation of the ECAPA-TDNN embeddings of the speech segments of a doctor-patient conversation marked with high or low confidence from the local confidence method (T3), including overlapping speech. We note that the speech segments farther from speaker centroids (possibly due to noise or high intra-speaker variance) are assessed as low-confidence segments. We also note that the proposed method assesses most overlapping speech segments as low-confidence segments, likely due to the speaker recognition system's inability to assign overlapping speech segments to either of the speakers. This demonstrates a possible limitation of the proposed method, as it will remove most overlapping speech segments when used to prune an overlap-aware speaker diarization system. However, in the future, we plan to use this limitation to prune outliers of an overlap speech detector~\cite{Bredin2021} and combine it with the currently proposed method to develop an overlap-aware confidence assessment and thresholding technique.

\section{Conclusion}
Speaker diarization systems face many challenges due to the highly volatile nature of multi-talker speech conversations captured in unconstrained environments. While it is essential to develop speaker diarization systems robust to such challenges, downstream tasks that rely on speaker identity may be able to mitigate diarization errors where the system provides informative confidence measures. We conducted experiments across multiple datasets and diarization systems, comparing several different methods for assigning segment-level confidence to diarization systems' outputs. We found that silhouette score was always either the best or a close-second-best method for confidence assessment. Nonetheless, the top three methods were all able to isolate $\sim$30\% of the diarization errors within segments with the lowest $\sim$10\% of confidence scores, and $\sim$55\% of the diarization errors within segments with the lowest $\sim$30\% of confidence scores, indicating their predictive value.

\bibliographystyle{IEEEbib}
\bibliography{strings,refs}

\atColsBreak{\vskip5pt}
\end{document}